\documentclass[epjST]{svjour}
\usepackage{graphics,graphicx}
\usepackage{amssymb}
\usepackage{amsmath}
\usepackage{bm}
\usepackage{color}
\begin{document}
\title{Finding quark content of neutron stars in light of GW170817}
\author{Rana Nandi\inst{1}\fnmsep\thanks{\email{nandi.rana@gmail.com}} \and Subrata  Pal\inst{2} }
\institute{Polba Mahavidyalaya, Hooghly, West Bengal 712148, India \and 
Department of Nuclear and Atomic Physics, Tata Institute of Fundamental Research, Mumbai 400005, India}
\abstract{
The detection of gravitational waves from GW170817 has provided
a new opportunity to constrain the equation of state (EOS) of neutron stars. 
 In this article, we investigate the possible existence of 
 quarks inside the neutron star core in the context of GW170817.
 The nucleon phase is treated within the relativistic nuclear mean-field approach where we have employed a fully comprehensive set of available
 models, and  the quark phase is described in the Bag model.
  We show that the nucleonic EOSs which are inconsistent with 
 the tidal deformability bound become consistent when phase
 transition to quark matter via Gibbs construction is allowed.
 We find that several nucleonic EOSs support the presence
 of pure quark matter core with a small mass not more than $0.17M_\odot$ 
 confined within a radius of 0.9 km.
 We also find that the strong correlation between tidal deformability 
and neutron star radii observed for pure nucleonic stars 
does persist even with a nucleon-quark phase transition and provides
an upper limit on the radius of $R_{1.4} \lesssim 12.9$ km for 
a $1.4M_\odot$ neutron star.
} 
\maketitle

\section{Introduction}
\label{intro}

Neutron stars (NS) are highly compact astrophysical
objects which are produced at the end of the life
cycles of massive stars
($8M_\odot\lesssim M \lesssim 25M_\odot$) via supernova
explosions. A NS can have mass between
$\sim 1-2M_\odot$, but with a rather small radius of only
between $10-15$ km. As a result, the density inside the 
star can be very high $\sim 10^{15}-10^{16}$ g/cm$^3$,
which is several times larger than the saturation
density ($\rho_0\sim 2.8\times10^{14}$) g/cm$^3$ of
nuclear matter  \cite{Glendenning2000}. The state of
the matter, i.e. the equation of state (EOS) and the
composition, is not known at such high densities as
laboratory experiments and {\it ab initio} calculations
can only provide description of nuclear matter at
around the saturation density. The high-density EOS
of NS  matter is thus highly uncertain and it is explored by
adopting different models \cite{Oertel:2016bki}. 

In order to reliably constrain the EOS one should rely on
astrophysical observations \cite{Lattimer:2006xb}. Given an EOS, the Tolman-Oppenheimer-Volkoff equations
provide an unique sequence of masses and radii for NS
with the sequence terminating at a maximum mass
$M_{\rm max}$. The value of $M_{\rm max}$ depends on
the stiffness of the EOS, i.e. how rapidly the pressure
increases with the energy density, and a stiffer EOS 
generates a larger maximum mass star. Of course, the larger
matter pressure in a stiffer EOS state also generates 
stars with larger radii. Thus measurements
of masses and radii of NSs can put significant
constraint on the EOS \cite{Lindblom92}.

The appearance of new degrees of freedom,
such as quarks inside the core of neutron star, 
would soften the overall EOS resulting in decrease of maximum mass and radius.
In fact, the deconfinement transition from hadron to quark-gluon phase, as predicted
in the theory of strong interactions $-$ quantum chromodynamics, has been already
observed at high temperature and small net-baryon density
in ultra-relativistic heavy ion collisions.
In contrast, the presence of quark matter inside the high density core of neutron stars
still remains a open question. By combining astrophysical observations 
of mass and radii of neutron stars with various theoretical models of strongly
interacting matter one can provide empirical constraints on the quark-matter
content inside stars.

The first major observational breakthrough in this
direction came with the precise measurement of masses
of two massive NS with masses of 
$(1.928\pm0.017)M_\odot$ 
\cite{Demorest:2010bx,Fonseca:2016tux} and 
($2.01\pm0.04M_\odot$) \cite{Antoniadis:2013pzd}. Very
recently another massive NS of mass 
$2.14^{+0.20}_{-0.18}$ within  $95.4\%$ credibility 
interval ($2.14^{+0.10}_{-0.09}$, within $68.3\%$ credibility interval) 
has been detected \cite{Cromartie:2019kug}. These measurements
will essentially exclude the soft EOSs for which 
$M_{\rm max}<1.97M_\odot$. In fact, to determine
the EOS uniquely one also requires precise measurements of
radius of stars. A few measurements have been performed for NS 
radii from quiescent low-mass X-ray binaries and from
the thermonuclear bursts of accreting NS
\cite{Guillot:2013wu,Ozel:2015fia,Ozel:2016oaf,Nattila:2017wtj}. Although these measurements are important,
but these are unable to impose significant constraint on the EOS
as the uncertainty is quite large of $\sim 11-29\%$. 
NASA's {\it Neutron Star Interior Composition Explorer (NICER)} instrument was installed on the International
Space Station on 2017 with the mission partly  
to measure the masses and radii of NS within
$\sim 5\%$ uncertainty. Recently NICER 
collaboration has estimated \cite{Riley2019,Miller2019}
the mass ($M=1.34_{-0.16}^{+0.15}M_\odot$) and radius
($R=12.71^{+1.14}_{-1.19}$ km) of the millisecond
pulsar PSR J0030+0451.

On August 2017, LIGO-Virgo Collaboration (LVC)
detected first ever gravitational waves from the binary
NS merger event GW170817
\cite{TheLIGOScientific:2017qsa}.
This historic detection has opened up a new avenue to
constrain the EOS at high densities. During the
inspiral phase of a binary NS merger the strong 
gravitational field of each star tidally deform the 
other leaving detectable imprint in the emitted 
gravitational wave signal \cite{Hinderer:2009ca}.
By analyzing the data of GW170817, LVC obtained an
upper bound on the tidal deformability of a 
$1.4M_\odot$ neutron star of $\Lambda_{1.4}\leq800$. Due to its 
strong sensitivity on the radius ($\Lambda\sim R^5$),
tidal deformability can put stringent constraint on the EOS. 
Subsequently, several studies were carried
out to constrain the EOS \cite{Fattoyev:2017jql,Annala:2017llu,Most:2018hfd,Nandi:2017rhy,Zhang:2018vrx}
by using the tidal deformability bound provided by GW170817.
These studies provided an upper bound on the radius of
a $1.4M_\odot$ neutron star of $R_{1.4}\lesssim 13.5-13.8$ km
\cite{Fattoyev:2017jql,Annala:2017llu,Most:2018hfd,Nandi:2017rhy,Zhang:2018vrx}.
Upper bounds on the maximum mass $M_{\rm max}\lesssim 2.2M_\odot$ were also obtained by several authors
by analyzing the data of gravitational wave signal
as well as the electromagnetic counterparts of GW170817  
\cite{Margalit:2017dij,Rezzolla:2017aly,Shibata:2017xdx}. 
Both these bounds imply that the EOS cannot be very stiff.
Later LVC improved their analysis of GW170817 data by assuming a common EOS 
for both the stars and improved waveform model and obtained $\Lambda_{1.4}=190_{-120}
^{+290}$, which translates to an more stringent upper
bound of $\Lambda_{1.4}\leq 580$ \cite{Abbott:2018exr}.

Recently, we performed an extensive analysis of the widely-used relativistic
mean-field (RMF) model EOSs using the
observational constraints on the maximum mass of neutron star and tidal 
deformability of GW170817 and also employing the latest bounds on the 
saturation properties of nuclear matter \cite{Nandi:2018ami}. 
We found that only 3 out of 269 RMF model EOSs are consistent with all the constraints. 
Using a few selected nucleonic EOSs and limited range of quark matter parameters we further showed
that if the phase transition from nucleonic matter to quark
matter via Gibbs construction is incorporated in the EOS at
higher density, several EOSs become consistent with
all the observational bounds \cite{Nandi:2017rhy,Nandi:2018ami}. 
In this article we shall make a comprehensive analysis of the properties
of the neutron star with a nucleon-quark first order phase transition.
For this purpose we shall employ all the available nuclear RMF models for the nucleon sector
and the Bag model for the quark sector where the Bag model parameters are allowed to encompass
the entire permissible range of the quark matter parameter space. We shall show that majority of the
pure nucleonic model EOSs, that are consistent with the neutron star maximum mass bound
of $M_{\rm max}\geq1.97M_\odot$, do not satisfy the tidal deformability bound
of $\Lambda_{1.4}\leq 580$ \cite{Abbott:2018exr}. Inclusion of a quark phase in 
the neutron star softens the overall EOS, and we find that these stars become consistent
with the tidal deformability bound for realistic values of Bag model parameter space.
We shall also show that pure quark matter, though of small mass, can exist in the core
of neutron stars.

The article is organized as follows. In section 2 we provide the details of EOS calculation 
for both the nucleonic phase and the quark phase.
In section 3 we present the results for the maximum mass and radii of pure nucleon stars and 
with nucleon-quark phase transition. We discuss the resulting implications on the composition 
and content of quark matter in light of maximum mass and tidal deformability constraints. 
Finally, in section 4 we conclude with a discussion.

\section{Set up}
\label{sec:setup}

In this section we discuss the construction of EOSs
for both the nucleonic matter and quark matter and the phase 
transition between them. We also discuss the  calculation of tidal deformability of neutron stars.

\subsection{Nucleonic EOS}
We construct the EOS of the nuclear matter containing
neutrons, protons, electrons and muons by adopting
RMF approach introduced by Walecka \cite{Walecka:1974qa} and refined over the years by many authors
\cite{Boguta:1977xi,Sugahara:1993wz,Serot:1997xg,Horowitz:2000xj,Dhiman:2007ck}. 
In this model the interactions between nucleons are 
described via the exchange of several mesons. The  most general form of the Lagrangian can be written as
\cite{Dutra:2014qga}:
\begin{eqnarray}
 {\cal L}=&&\sum_N\bar{\psi}_N\left[\gamma^\mu\left(i\partial_\mu   - g_{\omega}  \omega_\mu - \frac{1}{2}g_\rho\bm{\tau\cdot\rho_\mu}\right)
 -\left(m_N- g_{\sigma }\sigma -g_{\delta }\bm{\tau\cdot\delta}\right)\right]\psi_N \nonumber \\
  && + \frac{1}{2}\left(\partial_\mu\sigma\partial^\mu\sigma - m_\sigma^2\sigma^2\right)
  - \frac{\kappa}{3!}(g_\sigma\sigma)^3 - \frac{\lambda}{4!}(g_\sigma\sigma)^4 \nonumber \\
  && -\frac{1}{4}\omega_{\mu\nu}\omega^{\mu\nu}
  + \frac{1}{2}m_\omega^2\omega_\mu\omega^\mu
  + \frac{\zeta}{4!}(g_\omega^2\omega_\mu\omega^\mu)^2 \nonumber \\
  &&-\frac{1}{4}\bm{\rho_{\mu\nu}\cdot\rho^{\mu\nu}}
   + \frac{1}{2}m_\rho^2\bm{\rho_\mu\cdot\rho^\mu}
   + \frac{1}{2}\left( \partial^\mu\bm{\delta}\cdot\partial_\mu\bm{\delta}-m_\delta\bm{\delta^2}\right)
  \nonumber \\
  && +\, g_\sigma g_\omega^2\sigma\omega_\mu\omega^\mu\left(\alpha_1+\frac{1}{2}\alpha_1'\right) + g_\sigma g_\rho^2\sigma\bm{\rho_\mu\cdot\rho^\mu}\left(\alpha_2+\frac{1}{2}\alpha_2'\right)
  \nonumber \\
  && +\frac{1}{2}\alpha_3'g_\omega^2g_\rho^2\omega_\mu\omega^\mu\bm{\rho_\mu\cdot\rho^\mu}
   \label{eq:Ld}
\end{eqnarray}
where, $\psi_N$ is the isospin doublet of nucleons,
$\sigma$, $\omega$, $\rho$ and $\delta$ represent 
scalar-isoscalar, vector-isoscalar, vector-isovector
and scalar-isovector meson fields, respectively.
There are another class of RMF models, where the nucleon-meson couplings are not constants but
density-dependent \cite{Typel:1999yq,Typel:2009sy}
and they do not contain any self-coupling or
cross-coupling terms of mesons. 

Some of the parameters appearing in the Lagrangian
are determined by fitting to the known saturation
properties of nuclear matter such as binding energy
per nucleon, the saturation density, the symmetry energy ($J$), the incompressibility ($K$)
and the nucleon effective mass ($m^*$) \cite{Glendenning2000}.
Rest of the parameters are essentially free and can be
varied to match various nuclear and NS properties. 
For certain EOSs, the binding energies and charge radii of some
finite nuclei are also used to determine the
parameters \cite{Dhiman:2007ck,Typel:1999yq,Reinhard:1989zi}. 
Out of 269 RMF parameter sets only 67 are found \cite{Nandi:2018ami}
consistent with the latest experimental/empirical
bounds on the following saturation properties \cite{Oertel:2016bki}:
\begin{eqnarray}\label{eq:sat}
210\, \leq &\, K\, ({\rm MeV})\, & \leq 280 \nonumber \\
28 \,\leq &\,  J\, ({\rm MeV})\, & \leq 35 \nonumber\\
30 \,\leq &\,  L\, ({\rm MeV})\, & \leq 87 
\end{eqnarray}
A wider range than the generally accepted values for incompressibility, 
namely $K=248\pm8$ MeV \cite{Piekarewicz:2003br} or $K=240\pm20$ MeV \cite{Shlomo2006}, 
were used because of  their model dependence
 \cite{Oertel:2016bki}. In this article we consider 
 all the RMF parameter sets which satisfy the above bounds and also consistent with observational bound
 on the maximum mass i.e. $M_{\rm max}\geq1.97M_\odot$.

\subsection{Quark EOS}

 To construct the EOS of quark matter we adopt the 
 modified MIT Bag model that provides phenomenological description
of the quark phase. The grand potential is given by \cite{Nandi:2017rhy,Weissenborn:2011qu}:
 \begin{equation}
\Omega_{\rm QM} = \sum_i \Omega_i^0 + \frac{3\mu^4}{4\pi^2}(1-a_4)+B_{\rm eff}, \label{eq:qm}
\end{equation}
where $\Omega_i^0$ denotes the grand potentials of
non-interacting Fermi gases of up ($u$), down ($d$) and strange ($s$) quarks and electrons. The other two
terms in Eq. (3) correspond to the strong interaction
correction and the nonperturbative QCD effects which are accounted via two effective parameters $a_4$ and $B_{\rm eff}$, with 
$\mu(=\mu_u+\mu_d+\mu_s$) being the baryon chemical potential of quarks.

We consider the phase transition from the nucleonic 
matter to the quark matter via Gibbs construction 
\cite{Glendenning2000,Glendenning:1992vb}
which is characterized by the appearance of a mixed
phase of nucleonic and quark matter between the pure
nucleonic and pure quark phases.

\subsection{Tidal deformability}\label{sec:td}

At the initial stage of an inspiraling binary NS, the 
tidal effect on a star can be written at linear order
as \cite{Hinderer:2007mb}:
\begin{equation}
Q_{ij}= - \lambda \mathcal{E}_{ij},
\end{equation}
where $Q_{ij}$ represents the induced quadrupole
moment of the star and $\mathcal{E}_{ij}$ is assumed
to be the external static tidal field exerted by the partner. The parameter $\lambda$ is related to the dimensionless quadrupole tidal love number $k_2$ as ($G=c=1$):
 \begin{eqnarray}\label{eq:td}
 \lambda &=& \frac{2}{3} k_2 R^5 ,\nonumber\\
 \Lambda &=& \lambda/M^5,
 \end{eqnarray}
 where $\Lambda$ is the dimensionless tidal deformability.
 
We follow the framework developed by Hinderer and collaborators 
\cite{Hinderer:2009ca,Hinderer:2007mb} to calculate $k_2$ and subsequently $\Lambda$.
The value of $k_2$ depends on the EOS and lies in
the range $\simeq0.05-0.15$ \cite{Hinderer:2009ca}.
This quantity can be expressed in terms of $C=M/R$, the  compactness parameter as:
\begin{eqnarray}
 k_2 = && \frac{8C^2}{5}(1-2C)^2 \left[2+2C(y-1)-y\right]\nonumber \\
&& \times \Bigg\{2C\left[6-3y+3C(5y-8)\right]
+ 4C^3\left[13-11y+C(3y-2) + 2C^2(1+y)\right]\nonumber \\
&& + 3(1-2C)^2[2-y+2C(y-1)]{\rm ln}(1-2C)\Bigg\}^{-1},
\end{eqnarray}
where $y$ is defined as $y\equiv y(r)|_{r=R}$. The function $y(r)$ can be obtained by
solving the following first-order differential equation:
 \begin{equation}\label{eq:k2}
r\frac{dy}{dr} + y(r)^2 + y(r)e^{\lambda(r)}\left\{1+4\pi r^2\left[p(r)-\varepsilon(r)\right] \right\}+r^2Q(r)=0,
 \end{equation}
 with 
 \begin{equation}
  Q(r) = 4\pi e^{\lambda(r)}
  \left[5\varepsilon(r) + 9p(r) + \frac{\varepsilon(r) + p(r)}{dp/d\varepsilon} \right] -6 \frac{e^{\lambda(r)}}{r^2} 
  - \left(\frac{d\nu}{dr}\right)^2,
 \end{equation}
\begin{equation}
 e^{\lambda(r)} = \left[1-\frac{2m(r)}{r}\right],\qquad \frac{d\nu}{dr} = \frac{2}{r}\left[\frac{m(r)+4\pi p(r)r^3}{r-2m(r)}\right]
\end{equation}
and boundary condition $y(0)=2$. Given an EOS and
the central pressure $p(0)$, the Love number and the tidal deformability can be obtained by solving
Eq. (\ref{eq:k2}) together with the 
Tolman-Oppenheimer-Volkoff (TOV) equations \cite{Glendenning2000}:
\begin{eqnarray}
 \frac{dp}{dr} &=& -\frac{\left[p(r)+\varepsilon(r)\right]\left[m(r)+4\pi r^3p(r)\right]}{r[r-2m(r)]}\\
 m(r) &=& 4\pi \int_0^r \varepsilon(r)r^2\, dr
\end{eqnarray}

\section{Results and Discussions}
\label{sec:results}

We construct an EOS with nucleon-quark phase
transition via Gibbs construction. For the nucleonic
part we consider all the RMF EOSs which are
consistent with both the latest saturation
properties as given in Eq. (\ref{eq:sat}) and the observational
lower bound on maximum mass $M_{\rm max}\geq1.97M_\odot$.
In Table \ref{tab:h_eos} we list all the nucleonic
EOSs considered here along with their saturation
properties and maximum mass. 
Since the choice of the crustal EOS does not significantly
affect the NS observables \cite{Biswas:2019ifs}, we employ the Baym-Pethick-Sutherland (BPS) EOS \cite{Baym:1971pw}. The crust-core matching is modeled in
a thermodynamics consistent fashion by following Ref.
\cite{Fortin:2016hny}. Also tabulated are the values
of tidal deformabilities for a $1.4M_\odot$ NS calculated using Eq (\ref{eq:td}). It is seen
that only three EOSs namely HC, TW99 and NL$\rho$ are consistent with the tidal deformability bound $\Lambda_{1.4}\leq580$
as also found in Ref. \cite{Nandi:2018ami},

\begin{table}
\centering
 \caption{Various relativistic nuclear mean-field models and their nuclear matter saturation properties, 
namely incompressibility $K$, symmetry energy $J$ and its slope $L$. For these nuclear RMF models 
some important observational properties are presented, namely the maximum mass of neutron star 
$M_{\rm max}$, the radii $R_{1.4}$ and tidal deformability $\Lambda_{1.4}$ of a $1.4M_\odot$ mass neutron star.} 
 \label{tab:h_eos}
 \begin{tabular}{lcccccc}
  \hline\noalign{\smallskip}
  EOS & $K$(MeV) & $J$(MeV) & $L$(MeV) & $M_{\rm max}/M_\odot$ & $\Lambda_{1.4}$ & 
  $R_{1.4}$(km)\\
  \noalign{\smallskip}\hline\noalign{\smallskip}
   FSUGarnet \cite{Utama:2016tcl} & $229.5$ & $30.9$ & $51.0$ & $2.07$ & $638$ & $12.95$ \\
   HC \cite{Bunta:2003fm} & $231.9$ & $31.0$ & $58.5$ & $2.28$ & $440$ & $12.26$  \\
   DDME2 \cite{Lalazissis:2005de} & $250.9$ & $32.3$ & $51.3$ &$2.48$ & $705$ & $13.02$ \\
   DD2 \cite{Typel:2009sy} &  $242.7$ & $31.7$ & $55.0$ & $2.42$ & $684$ & $13.16$ \\
   TW99 \cite{Typel:1999yq} & $240.3$ & $32.8$ & $55.3$ & $2.08$ & $403$ & $12.29$ \\
   DDME1 \cite{Niksic:2002ri} & $244.7$ & $33.1$ & $55.5$ & $2.44$ & $674$ & $13.16$\\
   DD \cite{Typel:2005ba} & $240.0$ & $31.6$ & $56.0$  & $2.41$ & $679$ & $13.15$\\
   NL3$\sigma\rho6$ \cite{Pais:2016xiu}& $270.0$ & $31.5$ & $55.0$ & $2.75$ & $974$ & $13.78$ \\
   NL3$\sigma\rho5$ \cite{Pais:2016xiu}& $270.0$ & $32.3$ & $61.0$ & $2.75$ & $986$ & $13.83$ \\
   NL3$\sigma\rho4$ \cite{Pais:2016xiu}& $270.0$ & $33.0$ & $68.0$ & $2.75$ & $1002$ & $13.91$ \\
   NL3$\sigma\rho3$ \cite{Pais:2016xiu}& $270.0$ & $33.9$ & $76.0$ & $2.75$ & $1027$ & $14.01$ \\
   NL3v6 \cite{Horowitz:2002mb} & $271.6$ & $32.4$ & $61.1$ & $2.75$ & $948$ & $13.77$ \\
   NL3v5 \cite{Horowitz:2002mb} & $271.6$ & $33.2$ & $68.2$ & $2.75$ & $965$ & $13.84$ \\
   NL3v4 \cite{Horowitz:2002mb} & $271.6$ & $34.0$ & $77.0$ & $2.75$ & $992$ & $13.95$ \\
   NL3v3 \cite{Horowitz:2002mb} & $271.6$ & $34.5$ & $82.1$ & $2.74$ & $1012$ & $14.01$  \\  
   S271v6 \cite{Horowitz:2002mb} & $271.0$ & $32.7$ & $59.8$ & $2.35$ & $629$ & $13.05$\\
   S271v5 \cite{Horowitz:2002mb} & $271.0$ & $33.3$ & $65.4$ & $2.34$ & $643$ & $13.12$ \\
   S271v4 \cite{Horowitz:2002mb} & $271.0$ & $33.8$ & $71.8$ & $2.34$ & $663$ & $13.23$ \\
   S271v3 \cite{Horowitz:2002mb} & $271.0$ & $34.4$ & $78.9$ & $2.34$ & $694$ & $13.35$ \\
   S271v2 \cite{Horowitz:2002mb} & $271.0$ & $35.0$ & $86.9$ & $2.34$ & $742$ & $13.51$ \\    
   BSR1 \cite{Dhiman:2007ck} & $239.9$ & $31.0$ & $59.4$ & $2.47$ & $797$ & $13.42$\\
   BSR2 \cite{Dhiman:2007ck} & $239.9$ & $31.5$ & $62.0$ & $2.39$ & $751$ & $13.34$ \\
   BSR3 \cite{Dhiman:2007ck} & $230.6$ & $32.7$ & $70.5$ & $2.36$ & $751$ & $13.39$ \\
   BSR4 \cite{Dhiman:2007ck} & $238.6$ & $33.2$  & $73.2$ & $2.44$ & $790$ & $13.49$ \\  
   BSR5 \cite{Dhiman:2007ck} & $235.8$ & $34.5$ & $83.4$ & $2.48$ & $838$ & $13.67$ \\
   IOPB-I \cite{Kumar:2017wqp} & $222.7$ & $33.3$ & $63.6$ & $2.15$ & $688$ & $13.27$ \\     
   BKA22\cite{Agrawal:2010wg} & $225.2$ & $33.2$ & $78.8$ & $1.97$ & $667$ & $13.29$ \\   
   NL$\rho$ \cite{Liu:2001iz} & $240.8$ & $30.4$ & $84.6$ & $2.09$ & $571$ & $12.81$\\
  \noalign{\smallskip}\hline
 \end{tabular}
\end{table}
We generate a large number of quark matter EOSs
corresponding to different values of $B_{\rm eff}^{1/4}$
and $a_4$ given in Eq. (3). These EOSs are then combined with all the nucleonic
EOSs considered via the Gibbs construction. However, we 
discard EOSs for which the starting density of mixed
phase is smaller than the crust-core transition density.

\begin{figure}
\centering
  \includegraphics[width=0.7\columnwidth,angle=-90]{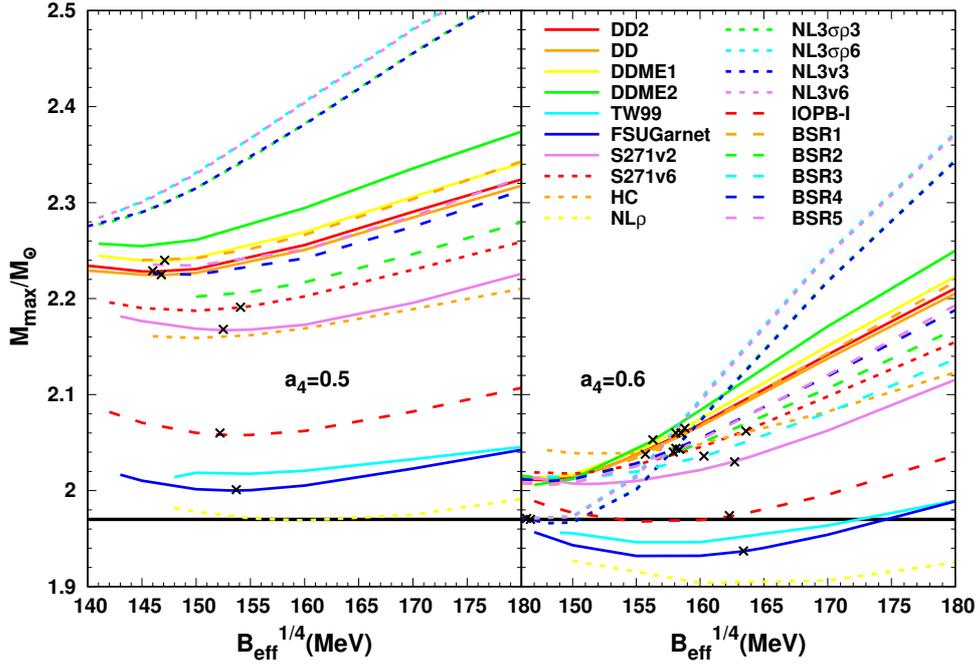} 
\caption{Maximum masses of neutron stars with nucleon-quark phase transition as a function of Bag pressure 
$B_{\rm eff}^{1/4}$ for $a_4=0.5$ (left panel) and $a_4=0.6$ (right panel) of Eq. (3) for various nucleonic 
EOSs as listed in Table \ref{tab:h_eos}. The black thick horizontal
line represents the lower bound $M_{\rm max}=1.97M_\odot$ on maximum mass. Crosses indicate maximum 
value of $B_{\rm eff}^{1/4}$ for stars that are consistent with $\Lambda_{1.4}\leq580$ bound.}
\label{fig:MmaxvsBeff}       
\end{figure}

Figure \ref{fig:MmaxvsBeff} shows the maximum masses
as a function of the Bag parameter $B_{\rm eff}^{1/4}$
for all the EOSs with a nucleon-quark phase transition obtained 
with values of $a_4=0.5$ (left panel) and $0.6$ (right panel). The
results for the EOS BKA22 are not shown as it gives a star with maximum mass of 
$1.97M_\odot$, and further addition of quarks makes the EOS softer leading to 
a $M_{\rm max}$ below the observed bound. 
Note that different EOSs within a
family are obtained by varying the single parameter,
namely $\alpha_3'$ for S271 and NL3v and $\alpha_2'$
\cite{Horowitz:2002mb,Pais:2016xiu}
for NL3$\sigma\rho$ (see Eq. (\ref{eq:Ld})) that provides 
different symmetry energy behavior without affecting
the $M_{\rm max}$, as can be seen from Table
\ref{tab:h_eos}. For each of these three families, 
we only display results corresponding to the highest and lowest values of the parameter;
the results for the other parameters fall in between these two limits.

At a fixed $a_4$, a small $B_{\rm eff}$ leads to a stiffer quark matter EOS as evident from
Eq. (\ref{eq:qm}) and noting that $P=-\Omega_{QM}$.
This causes the onset of phase transition i.e. the mixed
phase to occur early at a lower density and also of wider extent, resulting in softening of the
overall nucleon-quark EOS and generating star with smaller $M_{\rm max}$.
With increasing $B_{\rm eff}$, the quark phase has a smaller effect 
on the overall EOS due to its delayed appearance which causes the $M_{\rm max}$ to increase and eventually 
gives maximum mass for pure nucleonic star. Obviously, the effect is enhanced
for much stiffer quark matter EOS for large values of $a_4$.

It is evident from the Fig. \ref{fig:MmaxvsBeff}(right panel) that pure nucleonic EOS, viz  
FSUGarnet, TW99 and NL$\rho$ which have
maximum mass slightly above $2M_\odot$ (i.e. $M_{\rm max}<2.10M_\odot$)
cannot support stars with a maximum mass of $1.97M_\odot$ 
when quark phase is included with parameter value $a_4=0.6$. 
While nucleon-quark stars in FSUGarnet and TW99 fail to satisfy the maximum bound 
for smaller values of Bag parameter, the failure in NL$\rho$ EOS is for the entire range of 
$B_{\rm eff}^{1/4}$ studied here. All the other 17 EOSs for $a_4=0.6$ value are found
consistent with the maximum mass bound for the whole range of $B_{\rm eff}$.
In contrast, for $a_4=0.5$ (left panel), all the stars with nucleon-quark phase transition satisfy 
 the maximum mass bound.
By increasing the $a_4$ value to 0.6 and beyond 
causes more and more EOSs to fail the maximum mass constraint.
This is because the quark EOS becomes stiffer
with increasing $a_4$ as discussed above.

We now present results for the tidal deformability of neutron
stars with nucleon-quark phase transition following the
prescription presented in Sec. \ref{sec:td}. 
With increasing Bag constant, since the stiffer EOS 
generates stars with larger radii as well, the tidal deformability bound of 
will not be satisfied for large values of $B_{\rm eff}$. 
In Table \ref{tab:bag_max} we show the maximum
values of $B_{\rm eff}^{1/4}$ corresponding to different
nucleonic EOSs, for which the tidal deformability bound of 
$\Lambda_{1.4}\leq 580$ is satisfied by $1.4M_\odot$ neutron star
\cite{TheLIGOScientific:2017qsa}. 
These maximum values of $B_{\rm eff}^{1/4}$ are marked with ``crosses'' in 
Fig. \ref{fig:MmaxvsBeff}. The curves corresponding to
nucleonic EOS, DDME2, NL3$\sigma\rho3$, NL3$\sigma\rho6$, NL3v3,
NL3v6, BSR1-5 with $a_4=0.5$ and NL3$\sigma\rho6$ and NL3v6
with $a_4=0.6$ are not marked with any cross as these EOSs are unable
to satisfy the $\Lambda_{1.4}\leq 580$ bound
for any value of $B_{\rm eff}^{1/4}$. On the other hand TW99,
HC and NL$\rho$ EOSs satisfy the $\Lambda_{1.4}$ constraint for 
pure nucleonic stars. Since the inclusion of quarks makes 
the overall EOS softer resulting in stars with smaller masses and radii, the bounds are naturally satisfied 
for all values of $B_{\rm eff}^{1/4}$ and hence these EOSs with nucleon-quark phase are not marked with any cross.
It is interesting to note that out of the 17 
pure nucleonic EOSs that are not consistent with the 
tidal deformability constraint, 15 EOSs (except NL3$\sigma\rho6$ and NL3v6) for a range of values of
$B_{\rm eff}$ and $a_4$ can generate neutron stars with quark phase that are
consistent with the bound. However, the tidal
deformability bound is found to severely constrain
the quark matter parameter space ($B_{\rm eff}^{1/4},a_4$), irrespective of the nucleonic EOS.

\begin{table}\label{tab:bag_max}
 \centering
 \caption{ Listed for various nucleonic EOSs are the maximum values of $B_{\rm eff}^{1/4}$ 
that are consistent with the upper bound on $\Lambda_{1.4} \leq 580$ for the parameter values 
$a_4=0.5$ and 0.6 of Eq. (3). The corresponding radii $R_{1.4}$ of a $1.4M_\odot$ star are given.
NA denotes that No Allowed value of $B_{\rm eff}^{1/4}$ are consistent 
with the bound. The last column gives the maximum mass of the pure quark part (see text for details).}
  \begin{tabular}{lccccc}
 \hline\noalign{\smallskip}
  Hadronic & \multicolumn{2}{c}{$B_{\rm eff}^{1/4}|_{\rm max}$} & \multicolumn{2}{c}{$R_{1.4} (\rm km)$} & $(\Delta M_{\rm Q}/M_\odot )_{\rm max}$ \\
         EOS & $a_4=0.5$ & $a_4=0.6$ & $a_4=0.5$ & $a_4=0.6$    \\
 \noalign{\smallskip}\hline\noalign{\smallskip}
 FSUGarnet & 153 & 163 & 12.783 & 12.800 & 0.00 \\
 DDME2     & NA  & 156 & NA     & 12.863 & 0.02 \\
 DD2       & 146 & 158 & 12.821 & 12.892 & 0.01 \\
 DDME1     & 147 & 158 & 12.851 & 12.882 & 0.01 \\
 DD        & 146 & 158 & 12.808 & 12.878 & 0.01 \\
 NL3$\sigma\rho6$ & NA & NA & NA & NA & 0.13 \\
 NL3$\sigma\rho3$ & NA & 146 & NA & 12.518 & 0.17  \\
 NL3v6 & NA & NA & NA & NA & 0.14\\
 NL3v3 & NA & 146 & NA & 12.520 & 0.16\\
 S271v6 & 154 & 163 &12.923 & 12.916 & 0.00\\
 S271v2 & 152 & 162 & 13.049  & 13.064 & 0.00\\
 BSR1 & NA & 155 & NA & 12.835 & 0.02\\
 BSR2 & NA & 158 &  NA & 12.901 & 0.00 \\
 BSR3 & NA & 160 & NA & 12.966 & 0.00\\
 BSR4 & NA & 158 & NA & 12.967 & 0.02\\
 BSR5 & NA &157 & NA & 12.971 & 0.02\\
 IOPB-I & 152 & 162 & 12.979& 12.994 & 0.00 \\
  \noalign{\smallskip}\hline
 \end{tabular}
\end{table}

\begin{figure}
\centering
  \includegraphics[width=\columnwidth]{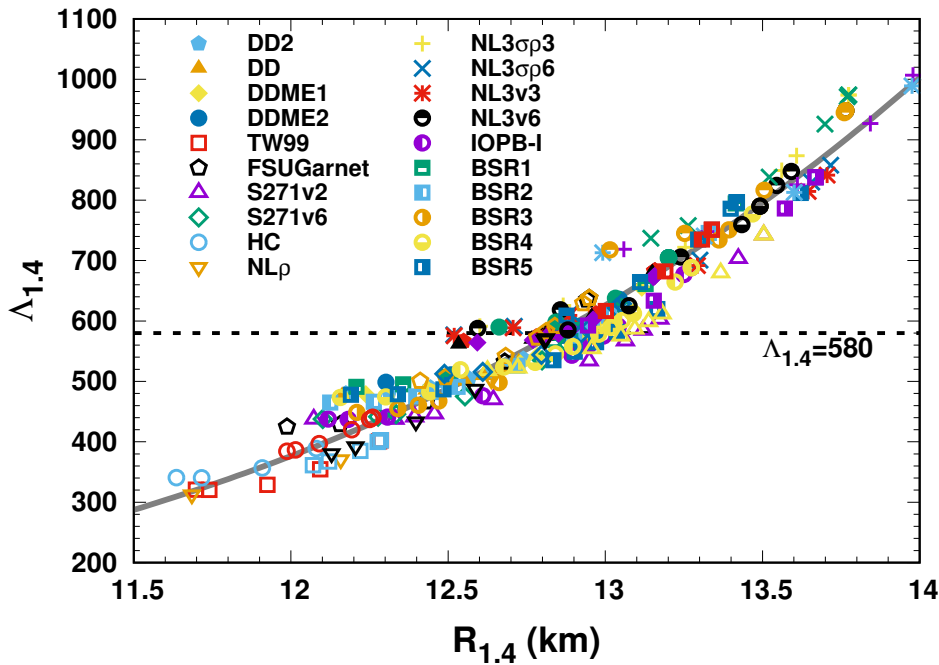} 
\caption{Correlation between $\Lambda_{1.4}$ and $R_{14}$ for EOS with nucleon-quark phase transition 
constructed from different nuclear EOSs and for a range of Bag parameter 
values $B_{\rm eff}^{1/4}\sim 145-180$ MeV and $a_4=0.5$ and 0.6. 
The dashed line represents the upper bound on $\Lambda_{1.4}$ given by GW170817 \cite{Abbott:2018exr} and
the solid line is for the fit $\Lambda_{1.4}=5.22\times(R_{14}/{\rm km})^{6.35}$.}
\label{fig:tdvsR} 
\end{figure}

The strong correlation between $\Lambda_{1.4}$ and $R_{1.4}$, as expected due to $\Lambda\propto R^5$, 
has been explored within various nuclear model approaches (without quarks)
\cite{Fattoyev:2017jql,Annala:2017llu,Tews:2019cap,De:2018uhw,Malik:2018zcf}. For all the RMF 
EOSs considered here and pure-nucleon stars, we obtained \cite{Nandi:2018ami}
the relation $\Lambda_{1.4}=1.53\times10^{-5}(R_{1.4}/{\rm km})^{6.83}$, with
maximum deviation $|(\Lambda_{1.4}^{\rm fit}-\Lambda_{1.4})/\Lambda_{1.4}|$ of $\sim 8\%$.
The extra factor of 1.83 in the exponent stems from the quadrupole love number $k_2$ which depends on the EOS 
and therefore on the radius of the star in a complicated fashion (see Sec. \ref{sec:td}).
In Fig. \ref{fig:tdvsR}, we present the correlation between $\Lambda_{1.4}$ and $R_{1.4}$ 
using these nucleonic EOSs and incorporating nucleon-quark phase transition.
We find that the strong $\Lambda_{1.4} - R_{1.4}$ correlation observed for pure nucleonic stars 
still persists with phase transition which 
can be fitted as $\Lambda_{1.4}=5.22\times10^{-5}(R_{1.4}/{\rm km})^{6.35}$. However, 
the correlations with quark phase have a slightly more spread, the maximum
deviation is $\sim 16\%$. Using this
fit function and the upper bound on $\Lambda_{1.4}\leq580$,
we obtain an approximate upper bound on the radius of  
$R_{1.4}\leq 12.9$ km. Interestingly, the same bound was obtained 
on $R_{1.4}$ for nucleon-only stars constructed from the RMF EOSs \cite{Nandi:2018ami}.

\begin{figure}
 \centering
 \includegraphics[width=0.7\columnwidth,angle=-90]{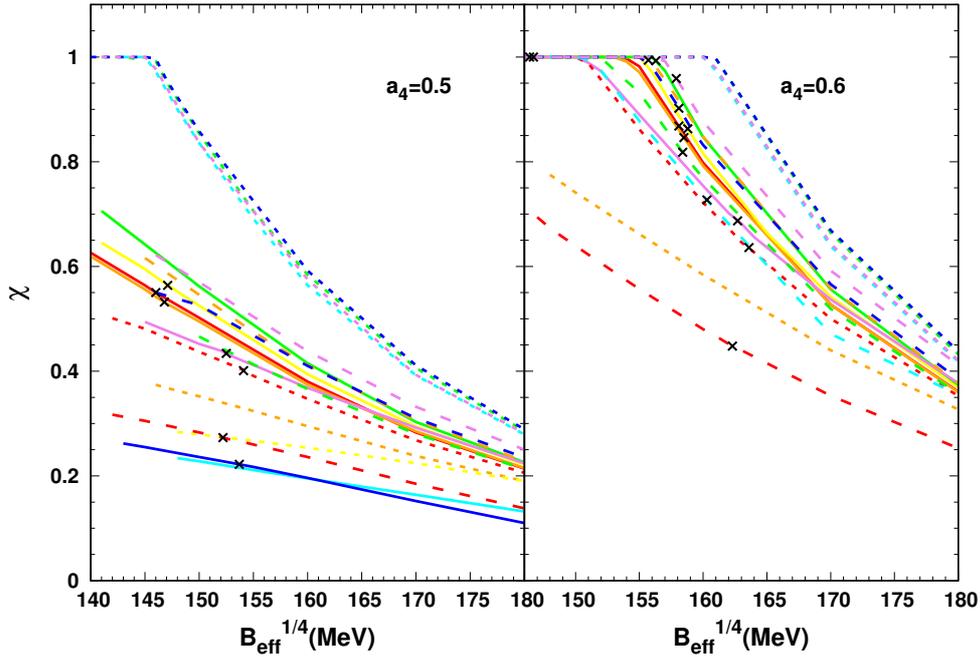}
 \caption{Volume fraction of quarks $\chi$ corresponding to maximum mass configuration as a 
function of $B_{\rm eff}^{1/4}$ for $a_4=0.5$ (left panel) and $a_4=0.6$ (right panel) 
for different nucleonic EOSs. 
The line styles and crosses are same as in Fig. \ref{fig:MmaxvsBeff} for the various EOS.}
\label{fig:qf_mmax}
\end{figure}

In Fig. \ref{fig:qf_mmax} we show the volume fraction $\chi$ of
quarks for the maximum mass configurations corresponding to
different Bag parameters $B_{\rm eff}^{1/4}$ and values of $a_4 =0.5$ and 0.6.
Only the EOSs of Fig. \ref{fig:MmaxvsBeff}, that satisfy the maximum mass bound 
for the quark matter parameters used, are considered here. The cross
indicate the maximum values of $B_{\rm eff}^{1/4}$ for which the corresponding 
EOS is consistent with the $\Lambda_{1.4}\leq 580$ bound. 
(NL3$\sigma\rho6$ and NL3v6 are not shown as these cannot support $\Lambda_{1.4}\leq580$.) 
We observe that with increasing $B_{\rm eff}$,
the fraction of quarks in the star decreases. This can be explained
from the fact that larger $B_{\rm eff}$ makes the quark EOS softer 
which delays the appearance of mixed phase to a higher density.
Consequently, the overall EOS becomes stiffer resulting in higher
maximum mass for a neutron star but at a lower central density
(see Fig. \ref{fig:MmaxvsBeff}). 

In the mixed phase, the quark fraction increases from $\chi=0$ (pure nucleonic phase) 
to $\chi=1$ (pure quark phase) as the density increase.
Since the maximum density inside the star is lower for a
higher $B_{\rm eff}$, the corresponding quark fraction is also smaller. 
Figure \ref{fig:qf_mmax} reveals that for $a_4=0.6$ there are several RMF models
for which the neutron star core can have pure quark matter 
while satisfying the $\Lambda_{1.4}\leq580$ constraint. Whereas,
for $a_4=0.5$, no such EOS exists that permits a pure quark matter core. Instead, 
the neutron star core consists of a mixed phase of nucleons and quarks.
In Table \ref{tab:bag_max} we have also listed
the maximum masses of the pure quark phase star $(\Delta M_{\rm Q})_{\rm max}$.
For a given RMF EOS, each combination of the parameters $(B_{\rm eff}^{1/4}, a_4)$
defines a value of $\Delta M_{\rm Q} = M_{\rm max}-M_{\rm mp}$, where $M_{\rm mp}$ is the mass of the 
star with the end point of the mixed phase as the central density. $(\Delta M_{\rm Q})_{\rm max}$ 
then corresponds to the maximum value of $\Delta M_{\rm Q}$ obtained by considering all possible combinations of 
$(B_{\rm eff}^{1/4}, a_4)$. From Table \ref{tab:bag_max}, we
find that NL3$\sigma\rho3$ and NL3v3 EOSs have appreciable 
size of quark-matter core with mass of $\sim 0.17 M_\odot$ which 
corresponds to $\sim 8\%$ of total mass of the star. For other
EOSs the quark core mass is quite small up to $0.02M_\odot$.

\begin{figure}
 \centering
 \includegraphics[width=\columnwidth]{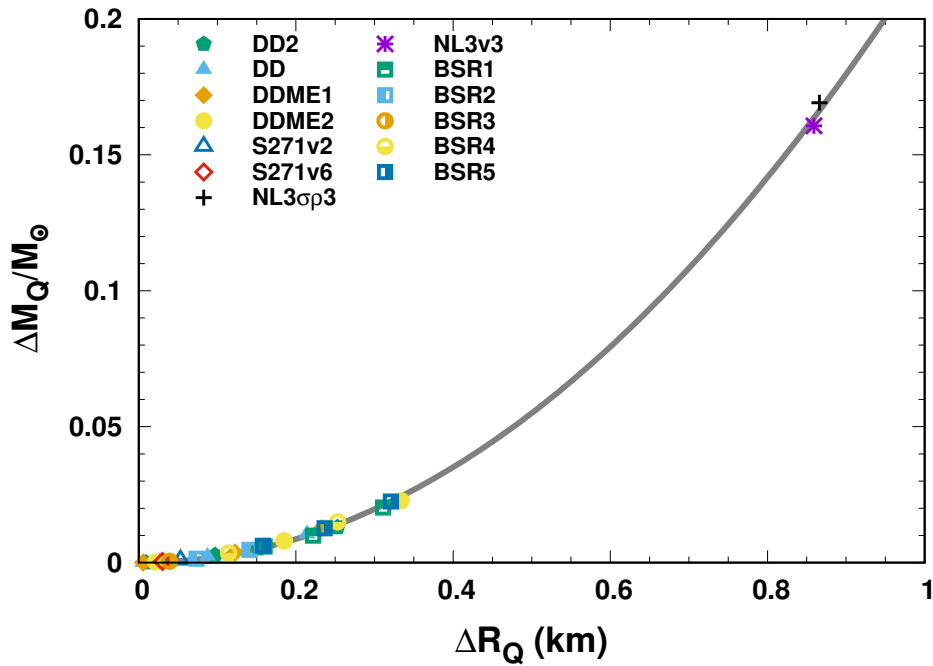}
 \caption{Masses and radii of quark cores for the maximum mass
 neutron star configurations corresponding to different nucleonic EOSs
 and Bag model parameters that are consistent with
 $M_{\rm  max}\geq1.97M_\odot$ and $\Lambda_{1.4}\leq580$ bounds. The thick grey line represents the fit 
 $\Delta M_Q/M_\odot=0.22\times(\Delta R_Q/{\rm km})^{2.01}$.}
\label{fig:qcore}
\end{figure}

In Fig. \ref{fig:qcore}, we display the variation of masses $\Delta M_Q$ with the 
radii $\Delta R_Q$ of the quark-matter core for the  maximum mass neutron star
configurations corresponding to different RMF nucleonic EOS
and Bag model parameters. We show only those
configurations which are consistent with both the maximum
mass bound and the tidal deformability bound. In the present model analysis, while 
the maximum mass of $0.17 M_\odot$ predicted for the quark-matter core is confined within a radius 0.9 km, 
the majority of the models lead to a much smaller masses and radii of $\sim 0.02M_\odot$ and $\sim 0.3$ km. 
It is interesting to observe that the mass and radius of the quark core are strongly correlated and can be fitted
as $\Delta M_Q/M_\odot=0.22\times(\Delta R_Q/{\rm km})^{2.01}$.
We note that the central densities of these quark core stars are found 
early in the pure quark phase, immediately after the mixed phase, instead at very high
densities where all the quark EOSs have the same speed of sound ($c_s^2 = 1/3$). 
However, we found that speed of sound for these quark core stars are nearly similar
which result in a strong correlation between $\Delta M_Q$ and $\Delta R_Q$. 
Nevertheless, it may be worth investigating where other
model approaches lead to such a tight correlation in $\Delta M_Q$ and $\Delta R_Q$.

\section{Conclusions}

Observation of $\sim 2M_\odot$ neutron stars and the measurement of
tidal deformability from GW170817 have posed serious challenge
to the construction of EOS of a neutron star. While the 
maximum mass bound enforces a stiff EOS, the tidal deformability
bound $\Lambda_{1.4}\leq 580$ demands a soft EOS. 
A natural way to account such a behavior is by incorporating a nucleon-quark phase 
transition in the EOS at higher densities. In this work
we have investigated this possibility by considering nucleonic EOS from several RMF models, that are compatible with
constraints imposed by experimental data and observations, and including a quark matter EOS (via Bag model)
by exploring a wide range of quark matter parameter space. The EOSs with 
phase transition are generated via Gibbs construction 
characterized by nucleon-quark mixed phase. We have shown that
most of the nuclear EOSs that do not satisfy the tidal deformability bound, become consistent 
with this bound when transition to quark-matter is included for a rather large combination of 
Bag model parameters ($B_{\rm eff}, a_4$). However, the tidal deformability constraint is found 
to significantly reduce the allowed region of quark matter parameter space, regardless of 
the nucleonic EOS. We find that, for most of the nucleonic models studied, the neutron star core contains
a mixed phase of nucleons and quarks. We also find that several EOSs can support a neutron star 
with a pure quark matter core, albeit with quite small quark core mass within the range of
$\sim (0.02 -0.17)M_\odot$. Furthermore, we showed that a 
strong correlation exists between the masses and radii of the quark matter core.

Apart from the three RMF nucleonic EOSs found in this study,
there are few other nucleonic EOSs (e.g. APR \cite{Akmal:1998cf}, SLy \cite{Douchin:2001sv}) 
which are consistent with the two solar mass and tidal deformability bounds.  
Therefore, it is quite difficult to distinguish purely nucleonic stars
from hybrid stars with small quark core and/or mixed phase,
observationally. Nevertheless, recent binary neutron star simulations \cite{Weih:2019xvw} 
have shown that in the so-called delayed phase transition scenario a
hyper-massive hybrid star can be formed. During this process, the emitted 
gravitational wave can provide signature of hybrid stars even with a mixed phase.
However, the signature is strong for stars with significant quark core. 
The present study may thus be quite promising in the search for hybrid stars.

\end{document}